\documentclass[twocolumn,amsmath,amssymb,aps,prl,reprint,longbibliography,floatfix,8pt]{revtex4-1}
\usepackage{cancel}
\usepackage{enumitem}
\usepackage[utf8x]{inputenc}
\usepackage{graphicx}
\usepackage{dcolumn}
\usepackage{bm}
\usepackage{hyperref}
\usepackage[switch]{lineno}
\usepackage{float}             
\usepackage{subfigure}
\usepackage{tikz}
\usetikzlibrary{arrows.meta}

\begin{document}

\title{A complete ab initio view of Orbach and Raman spin-lattice relaxation in a Dysprosium coordination compound}

\author{Matteo Briganti}
\author{Fabio Santanni}
\author{Lorenzo Tesi}
\author{Federico Totti}
\author{Roberta Sessoli}
\email{roberta.sessoli@unifi.it}
\affiliation{Department of Chemistry "Ugo Schiff", INSTM Research Unit, Università degli Studi di Firenze, Sesto F.no, Italy}
\author{Alessandro Lunghi}
\email{lunghia@tcd.ie}
\affiliation{School of Physics, AMBER and CRANN Institute, Trinity College, Dublin 2, Ireland}

\begin{abstract}
{\bf The unique electronic and magnetic properties of Lanthanides molecular complexes place them at the forefront of the race towards high-temperature single-ion magnets and magnetic quantum bits. The design of compounds of this class has so far been almost exclusively driven by static crystal field considerations, with emphasis on increasing the magnetic anisotropy barrier. This guideline has now reached its maximum potential and new progress can only come from a deeper understanding of spin-phonon relaxation mechanisms. In this work we compute relaxation times fully \textit{ab initio} and unveil the nature of all spin-phonon relaxation mechanisms, namely Orbach and Raman pathways, in a prototypical Dy single-ion magnet. Computational predictions are in agreement with the experimental determination of spin relaxation time and crystal field anisotropy, and show that Raman relaxation, dominating at low temperature, is triggered by low-energy phonons and little affected by further engineering of crystal field axiality. A comprehensive analysis of spin-phonon coupling mechanism reveals that molecular vibrations beyond the ion's first coordination shell can also assume a prominent role in spin relaxation through an electrostatic polarization effect. Therefore, this work shows the way forward in the field by delivering a novel and complete set of chemically-sound design rules tackling every aspect of spin relaxation at any temperature.}
\end{abstract}

\maketitle

\section*{Introduction}

Contrary to what the name rare-earths can suggest, lanthanide elements (Ln) find a widespread application in numerous technological\cite{DeBettencourt-Dias2014} and bio-medical applications\cite{cotruvo2019}. The partially filled f-orbital shell and the large spin-orbit coupling interaction grant their compounds unique magnetic and electronic properties. For instance, they are employed for the mass fabrication of permanent magnets\cite{Coey2002}, while several Ln ions make for ideal constituent of luminescent probes\cite{bunzli2015,bunzli2019}, sensors\cite{cable2011}, LEDs\cite{zhou2018}, and MRI contrast agents\cite{Caravan1999, bottrill2006}. Over the last two decades, coordination compounds of trivalent Ln ions also acquired a prominent role in molecular magnetism since the discovery that they can show slow relaxation of the magnetization and the opening up of a hysteresis loop\cite{winpenny2013, Liddle2015,Liu2018,Coronado2020}, offering a glance at the possibility to create a single-molecule version of bulk permanent magnets for high-density memory storage. Interestingly, the possibility of obtaining Ln complexes with very long magnetic moment lifetimes\cite{shiddiq2016, Gaita-Arino2019} also makes them optimal candidates for the realization of multi-level quantum bits (qudits)\cite{Godfrin2017,Hussain2018,Moreno-Pineda2018}.

All these applications have one common denominator: their working mechanism stems from a very efficient decoupling of the Ln's electronic and magnetic degrees of freedom from the atomic motion of the ion's environment.  The latter, acting as a thermal bath for the electronic degrees of freedom, invariably leads to electronic and spin relaxation and the quenching of the desired properties. In the case of magnetic relaxation of Kramers ions, this decoupling is realized by preventing an electronic state with maximum $z$-component of the total angular momentum, namely $m_J=J$, to invert its orientation to $m_J=-J$. This can be accomplished by a suitable choice of the Ln ion's crystal field in a way that the Kramers' doublet (KD) $m_J=\pm J$ is the ground state and the other states with smaller projection of $m_J$ are well separated in energy and only slightly admixed among them\cite{Sorace2011,Ungur2016}. In such a condition the transition within the ground state KD can only be mediated by a sequential number of transitions involving excited KDs and induced by a sequential absorption/emission of one quantum of lattice vibrational energy\cite{Liddle2015}, namely a phonon. Ideally, the system would need to climb up to the highest KD, whose energy corresponds to the maximum magnetic anisotropy of the system. This relaxation mechanism, called the Orbach process, becomes very slow at low temperature due to the lack of thermally available phonons and it follows an exponential relation of the type $\tau=\tau_0 exp(U/k_{B}T)$, where the pre-exponential factor $\tau_0$ sets the time scale of relaxation $\tau$, and $U$ is the anisotropy barrier that the magnetic moment needs to overcome in order to invert its orientation. This temperature dependence has been successfully used to reproduce the behavior of 3d SMMs such as Mn$_{12}$, Fe$_{8}$, etc.\cite{Sessoli1993,Sessoli,Pontillon1999}, 
whose spin levels span an energy range smaller than the energy of the first optical vibrational modes. The increase of $U$ has been at the centre of the synthetic efforts in the field and has led to the design of molecular complexes with anisotropy barriers exceeding 2000 K\cite{Goodwin2017,Guo2017,Guo2018}. 

Although very effective synthetic strategies have pushed $U$ to its maximum potential value\cite{Escalera-Moreno2019}, their success is undercut by two adverse effects: the small value of $\tau_0$ and the rise of an additional relaxation mechanism at low temperature, generally referred to as Raman\cite{Liddle2015} on the basis of phenomenological models. 

Up to now, a clear chemical interpretation of $\tau_{0}$ and Raman relaxation in terms of molecular motion is not available for Ln complexes. Further progress in this field is only possible through understanding of the physical nature of spin-phonon relaxation by in-depth first-principles description of how the solid-state molecular environment affects the spin degrees of freedom. Only in the last few years significant advances have been achieved with the introduction of \textit{ab initio} methods for the prediction of spin-phonon coupling\cite{Lunghi2017a, Escalera-Moreno2017} and spin-phonon relaxation time\cite{Lunghi2017}. On the basis of these seminal contributions, both transition metal-based spin qubits \cite{Lunghi2019, Lunghi2020a, Garlatti2020, Mirzoyan2020} and mononuclear single-molecule magnets\cite{Rechkemmer2016, Lunghi2017, Moseley2018, Ullah2019, Lunghi2020, Gu2020, Escalera-Moreno2020, Gu2021} have been studied to disentangle the influence of the spin-vibrational coupling and of the effective phonon density of states on the temperature dependence of relaxation rates. The community has recently showed a large interest in simulating relaxation time in Ln complexes, including second-order mechanisms responsible for Raman relaxation\cite{Chiesa2020}, but so far investigations have either included phenomenological parameters\cite{Chiesa2020}, or have been limited to computing vibrations at the gas-phase molecular level\cite{Goodwin2017,Guo2018,Ullah2019,Reta2021}. Such approaches, although insightful, risk to hide important details of spin dynamics or do not take into account the lower part of the phonons spectrum. The latter has indeed been proposed to be crucial to understand low-temperature relaxation  mechanisms.\cite{Lunghi2017,Lunghi2020,Gu2020,Chiesa2020} 

In this contribution we make a significant step forward by providing the first accurate and fully-\textit{ab initio} description of the one- and two-phonon processes that lead to magnetic relaxation in a prototypical Ln single-ion magnet. The success of a fully \textit{ab initio} theory of magnetic relaxation is pivotal towards the unambiguous interpretation of the relaxation mechanisms and provides the stepping stone for a detailed analysis of the origin of relaxation itself. Backed up by a rigorous theoretical and computational framework, it is now possible to go beyond simple arguments based on phenomenological models, look at the molecular origin of spin-phonon coupling, and propose a comprehensive chemical strategy against fast spin-phonon relaxation.

The selected molecule for our study is [Dy(acac)$_3$(H$_2$O)$_2$]·EtOH·H$_2$O compound\cite{Jiang2010} (acac$^{-}$ = acetylacetonate, EtOH=ethanol), \textbf{Dyacac} for short. This molecule was among the first single-ion Ln complexes to have been reported to show slow relaxation of its magnetic moment in absence of a static external magnetic field and its magnetic relaxation behaviour represents a fingerprint for most of the many Ln SMMs reported in the literature. An additional factor in favor of \textbf{Dyacac} is that this compound is simple enough to permit a detailed analysis and, at the same time,  it allows a fully \textit{ab initio} investigation without prohibitive computational requirements.

Computational predictions of spin-relaxation time compare well with susceptometry measurements and allow to demonstrate that second-order Raman relaxation is indeed an excellent candidate to explain the low-temperature relaxation regime. Moreover, we perform a detailed investigation on how spin-phonon coupling emerges from molecular motion, thus providing a chemical interpretation of both Orbach and Raman relaxation. Surprisingly, our results show remarkably large contributions from atoms beyond the first coordination shell as a result of electrostatic polarization. This finding defies the intuition that 4f magnetic states, being strongly localized, are only affected by the distortions of the first coordination shell and point toward ligands with donor atoms weakly polarizable by intra-molecular vibrations as a new chemical strategy to reduce spin-phonon coupling. 

\section*{The Dy$^{3+}$ Crystal Field: Static \textit{Ab Initio} calculations and validation by Cantilever Torque Magnetometry}
In the investigated compound, the Dysprosium atom is bonded to three acac$^{-}$ ligands and two water molecules, with a total coordination number of eight. Although the geometry around the Ln ion can be approximated to a structure with point-group symmetry $D2d$, the actual symmetry is too low to provide hints about the orientation and the values of the components of the magnetic anisotropy tensor. \textit{Ab initio} simulations and cantilever magnetometry are used to infer the electronic structure of \textbf{Dyacac}.

In order to accurately simulate the chemical environment of the Dy ion, our model include i) the molecular unit made of [Dy(acac)$_3$(H$_2$O)$_2$], and two co-crystallized ethanol and water molecules; ii) a surrounding 7x5x7 supercell of point charges placed at the atomic crystallographic positions to reproduce the Madelung potential inside the crystal. The optimum size of the supercell was evaluated by computing the electronic structure as a function of the number of charges (see SI). All the calculations were performed at the Complete Active Space Self Consistent Field (CASSCF) level of theory, followed by Complete Active Space State Interaction by Spin-Orbit interaction (CASSI-SO) (see Methods in SI for further details).   

\begin{figure}
 \begin{center}
    \includegraphics[scale=0.3]{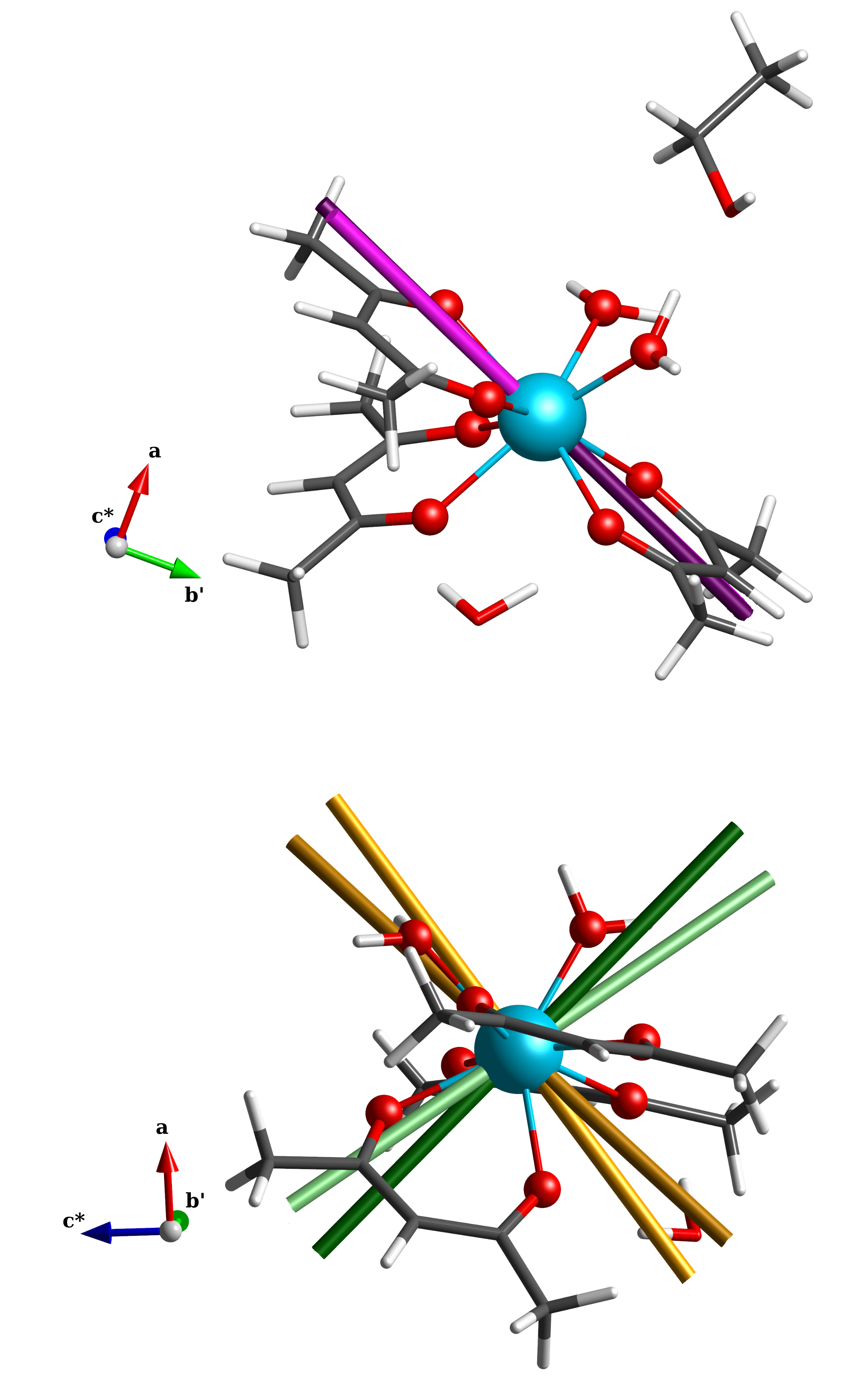}
\end{center}
 \caption{\textbf{Magnetization principal axis.}
  The molecular structure of \textbf{Dyacac} together with the principal anisotropy axes of ground KD as obtained by simulation with the \textit{ab initio} parameters (light colors) and by fit of CTM experimental curves (dark colors). Axis color code: magenta=easy axis, orange=hard axis and green=intermediate axis. Atomic color code: aquamarine=Dy, red= oxygen, grey=carbon, white=hydrogen.}
 \label{CTM2}
\end{figure}

Despite the lack of symmetry, the computed \textit{g}-tensor of the ground KD is quite axial, see Table S6, as is commonly found for Dy$^{3+}$ complexes despite the low symmetry coordination geometries, as recently explained.\cite{Briganti2021} The direction of the easy axis (see Fig. \ref{CTM2}) is almost parallel to the bond with the closest oxygen atom of one acac$^{−}$ ligand, as commonly found\cite{Chilton2013, Ding2020}. Such a direction can be rationalized in the following way: in order to minimize the electrostatic repulsion with the ligands, the oblate electron density of the $m_{J}=\pm 15/2$ doublet is mostly localized in the plane defined by the two water molecules' oxygens (neutral), which also present the largest bond lengths (see Fig. \ref{CTM2}).  

The computed electronic structure shows the first and second excited doublets lying at 101 and 129 cm$^{-1}$, respectively, and with the whole ${}^6H_{\frac{15}{2}}$ multiplet spanning 450 cm$^{-1}$. Such values are quite common for Dy complexes even in low symmetry environments, as demonstrated by the many examples existing in literature.\cite{Liddle2015} The accuracy of \textit{ab initio} calculations is validated by comparison with the experimental magnetic anisotropy determined by Cantilever Torque Magnetometry (CTM)\cite{Perfetti2015a,Lucaccini2016,Sorensen2018a,Mihalcea2016,Rigamonti2014,Perfetti2019a}. In particular, CTM on single crystals is a very sensitive tool allowing measurements over a wide temperature range, thus probing also excited spin states\cite{Perfetti2017, Rigamonti2018, Pointillart2020}.

\begin{figure}
 \begin{center}
    \includegraphics[scale=1]{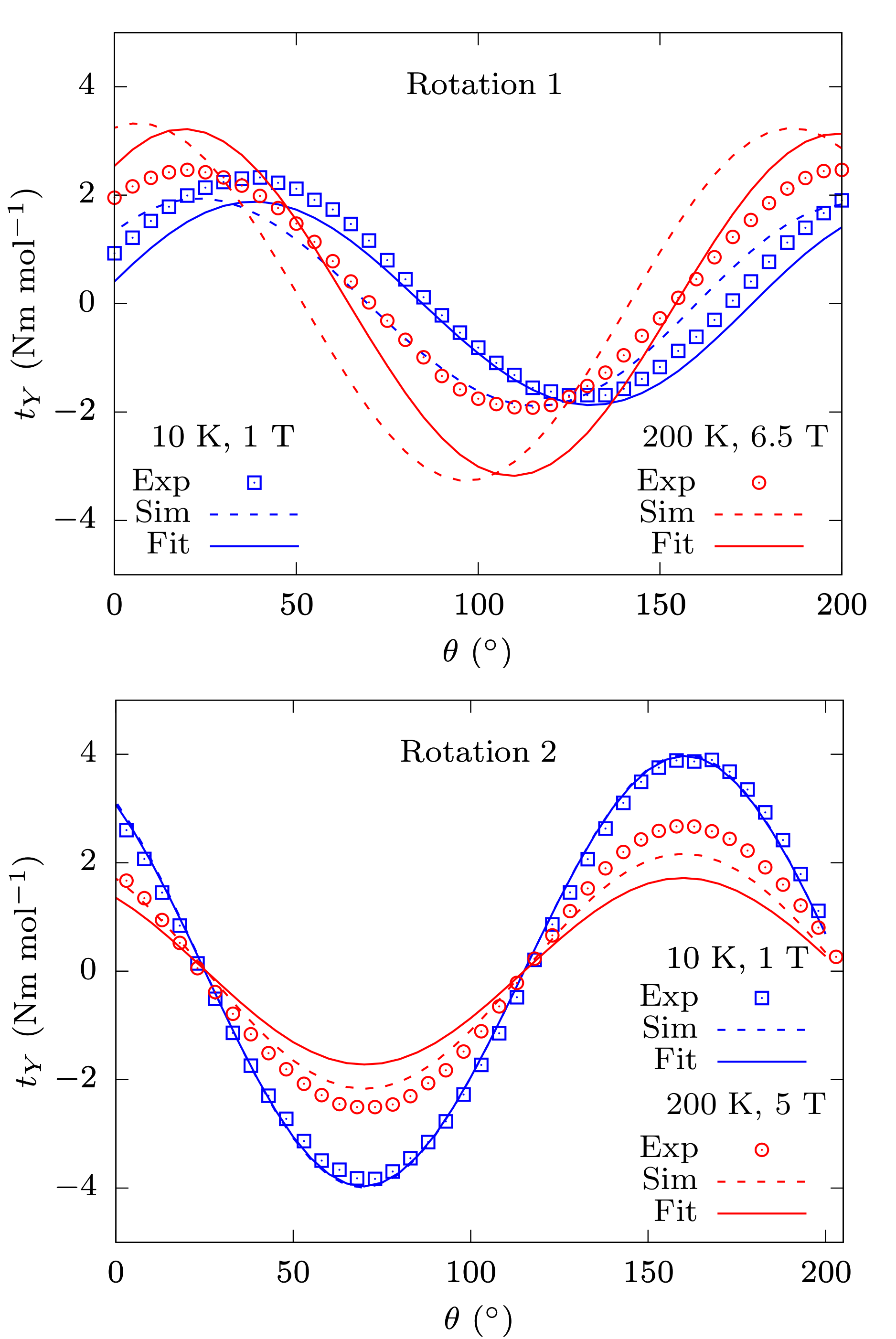}
\end{center}
 \caption{\textbf{Magnetization torque measurement.}
 The torque momentum measured on \textbf{Dyacac} single crystal as a function of the rotation angle at 10 K (blue color and empty square symbols) and 200 K (red color and empty square symbols) for Rot1 (top left panel) and Rot2 (bottom left panel). The simulations are based on the parameters obtained by \textit{ab initio} calculations. The fits are implemented by using the \textit{ab initio} crystal field parameters and rotating the magnetic tensor (see Methods in SI).}
 \label{CTM}
\end{figure}

 In the experiment, a freshly prepared \textbf{Dyacac} single crystal (see Methods in SI) was indexed by X-ray diffraction and placed on a rotating cantilever, which was inserted into the cryostat. Two sets of orthogonal rotations (Rot1 and Rot2) were performed under an external magnetic field, $\vec{B}$, with defined direction. The measured component of torque momentum $\vec{t}$ is given by equation:
\begin{equation}
    	t_Y (\theta)=M_Z B_X - M_X B_Z
    	\label{CTM_eq}
\end{equation}
where $XYZ$ is the laboratory reference frame, $\theta$ the rotation angle and $M$ the sample magnetization. By rotating the sample, it is possible to map the projection of the magnetic principal axes on a defined plane, and every time one of this is parallel to  $\vec{B}$, the torque is zero. Being the orientation of the crystallographic frame, $abc$*, known with respect of the laboratory reference frame, it is possible to identify the directions of the principal magnetic anisotropy axes in the molecular frame, $xyz$ (see Experimental Methods in SI). The results of the CTM experiments for the two rotations obtained for the lowest and highest investigated temperature values are shown in Fig. \ref{CTM}, while all results are displayed in Fig. S15-S16. 

\textit{Ab initio} calculations provide all information necessary to simulate the CTM experimental results, \textit{i.e.} the parameters $B^l_{m}$ of the crystal field Hamiltonian
\begin{equation}
    \hat{H}_{CF}=\sum_{l=2}^{14}\sum_{m=-l}^{l} B_m^l \hat{O}^l_m(\vec{\mathbf{J}})\:,
    \label{HCF}
\end{equation}
where $\hat{O}^l_m(\vec{\mathbf{J}})$ are tesseral tensor operators \cite{Tennant2000}.
These quantities allow to calculate magnetization in all directions of space and the magnetic anisotropy tensor at all temperatures. The comparison between experimental results and simulations is reported in Fig. \ref{CTM}. For Rot1, the simulations reproduce the experimental trend, despite a marked offset of ca. 14°. The Simulation overestimates the torque at high temperature. On the other hand, Rot2 exhibits a striking agreement between simulations and experiments and, remarkably, this agreement is observed even up to 200 K. The principal ground KD anisotropy axes obtained by \textit{ab initio} calculations can be visualized in Fig. \ref{CTM2} and Figs. S17-S18 in the SI. A refinement of the magnetic anisotropy orientation can be done by fitting the torque experimental data. As a consequence of the large number of parameters, the \textit{ab initio} crystal field parameters were kept fixed to the computed values, while the crystal field reference frame was left free to rotate. As it can be seen in Fig. \ref{CTM}, this procedure does not increase the already good agreement of simulations for Rot2, whereas it improves for Rot1. The angle among easy axes identified by simulations and fits is 10° (see Table S7), while a slightly larger gap can be found for the two axes with hard character (Fig. \ref{CTM2} and Figs. S17-S18). From this analysis, it emerges the high quality of \textit{ab initio} calculations, despite a minor tilt of the hard directions that are more difficult to reproduce.\cite{Pedersen2014, Marx2014, Ungur2017}

\section*{Spin-Lattice Relaxation}

The Crystal Field Hamiltonian of Eq. \ref{HCF} describes the energy levels of the ground state electronic multiplet for the equilibrium molecular geometry. However, at finite temperature, the position of the atoms inside the lattice fluctuates due to thermal energy. These structural fluctuations are the origin of spin-phonon coupling and have the ability to elicit transitions between different electronic states until the electronic energy levels reach the thermal equilibrium with the lattice. This phenomenon is captured by two sets of quantities, the vibrational modes of the lattice, $q_{\alpha}$, and the spin-phonon coupling operator $\hat{V}_{\alpha}=\sum_{lm}(\partial B_m^l / \partial q_\alpha )\hat{O}^l_{m}$, which describes the effects of lattice vibrations on $\hat{H}_{CF}$. All these parameters can be calculated from first principles following a strategy outlined on a series of previous works by some of the authors \cite{Lunghi2017, Lunghi2017a} and extensively benchmarked on transition metal ion qubits\cite{Lunghi2019, Albino2019, Lunghi2020a} and single-ion magnets \cite{Lunghi2020}. Details of all calculations are reported in the Methods section in SI. In a nutshell, phonons are calculated from periodic Density Functional Theory employing the crystal unit cell. \textbf{Dyacac}'s unit cell contains four molecular units and several interstitial water and ethanol molecules, for a total of 244 atoms. Vibrations computed on the periodic crystal's unit cell include low-frequency lattice distortions, where intra-molecular motion is admixed to rigid molecular translations and rotations. Including these contributions is crucial to fully capture the relaxation behaviour of these compounds and including all the phonons at the $\Gamma$-point is the first crucial step towards a full integration of the Brillouin zone as attempted elsewhere \cite{Lunghi2020}. Spin-phonon coupling coefficients are instead calculated by numerically differentiating the CF parameters by computing the values of $B^{l}_{m}$ along many molecular distortions with the CASSCF method. Once all these quantities have been computed, it is possible to predict the transition probability due to absorption or emission of phonons through a formalism very similar to the Fermi Golden Rule employed to predict spectroscopic transitions. In this work we consider first- and second-order time-dependent perturbation theory, where spin-phonon coupling, $V_{\alpha}$, is the perturbation and $\hat{H}_{CF}$ of Eq. \ref{HCF} is the unperturbed Hamiltonian. In this framework, it is possible to compute both one and two-phonon processes. In the nomenclature of spin-lattice relaxation, these processes correspond to the Orbach and Raman relaxation mechanism, respectively. The former relaxation mechanism involves the transition between two spin states $a$ and $b$ due to the absorption or emission of a single phonon $q_{\alpha}$. The rate of this process is described by Eq. \ref{order1}
\begin{equation}
W^\mathrm{1-ph}_{ba} = \frac{2\pi}{\hbar^{2}} \sum_{\alpha} \big|\langle b | \hat{V}_{\alpha} | a \rangle \big|^{2} G^\mathrm{1-ph}(\omega_{ba},\omega_{\alpha}) \:,
\label{order1}
\end{equation}
where $G^\mathrm{1-ph}=\delta(\omega-\omega_{\alpha})\bar{n}_{\alpha}+\delta(\omega+\omega_{\alpha})(\bar{n}_{\alpha}+1)$ and $\bar{n}_{\alpha}=[exp(\hbar\omega_{\alpha}/k_\mathrm{B}T)-1]^{-1}$ is the Bose-Einstein distribution of thermal population, $\hbar\omega_{\alpha}$ is the $\alpha$-phonon energy and $k_B$ is the Boltzmann constant. A similar but more rigorous expression for one-phonon process that includes the contribution of spin-states coherence was previously derived on the basis of the Redfield formalism\cite{Lunghi2019} (see Methods section in SI) and it is here used for the computation of one-phonon contributions to relaxation time.

The Raman relaxation mechanism accounts for transitions among two spin levels $a$ and $b$, and it is mediated by the simultaneous absorption and/or emission of two phonons and a contribution of all the electronic excited states. The Raman rate is modelled with the expression
\begin{align}
W^\mathrm{2-ph}_{ba} = \frac{2\pi}{\hbar^{2}} \sum_{\alpha\beta} & \Big|\sum_{c}\frac{\langle b |\hat{V}_{\alpha}|c\rangle\langle c|\hat{V}_{\beta}| a \rangle}{E_{c}-E_{a}\pm\hbar\omega_{\beta}}\Big|^{2}\cdot \label{order2} \\
& \cdot G_{\pm}^\mathrm{2-ph}(\omega_{ba},\omega_{\alpha},\omega_{\beta}) \nonumber \:,
\end{align}
where $G_{\pm}^{2-ph}$, reported in full in the Methods sections in SI, accounts for the thermal populations of phonons and imposes the conservation of energy, analogously to $G^{1-ph}$ for the Orbach process. There are three possible processes involving two phonons: absorption of two phonons, emission of two phonons and simultaneous emission of one phonon and absorption of another one. For instance, the emission of a phonon $q_{\alpha}$ and the absorption of a phonon $q_{\beta}$ contribute to $G_{\pm}^{2-ph}$ as 

\begin{equation}
    G^{2-ph}_{-}(\omega_{ba},\omega_{\alpha},\omega_{\beta}) = \delta(\omega_{ba}-\omega_{\beta}+\omega_{\alpha})\bar{n}_{\beta}(\bar{n}_{\alpha}+1)\:.
    \label{Gsph}
\end{equation}

The Raman process connecting two spin states $a$ and $b$ also involves a contribution from all the other spin states $c$. The contribution of this envelope of intermediate excited states, often referred to as a virtual state, does not involve any intermediate real transition to any of those states $c$, as it instead happens for a series of one-phonon processes in the Orbach relaxation. On the contrary, it only represents the fact that the spin states are no longer eigenstates of the total system (spin plus phonons) and that in such a condition all the KDs get slightly admixed among them by the external perturbation, \textit{i.e.} the phonons. 

\begin{figure*}
 \begin{center}
  \includegraphics[scale=0.35]{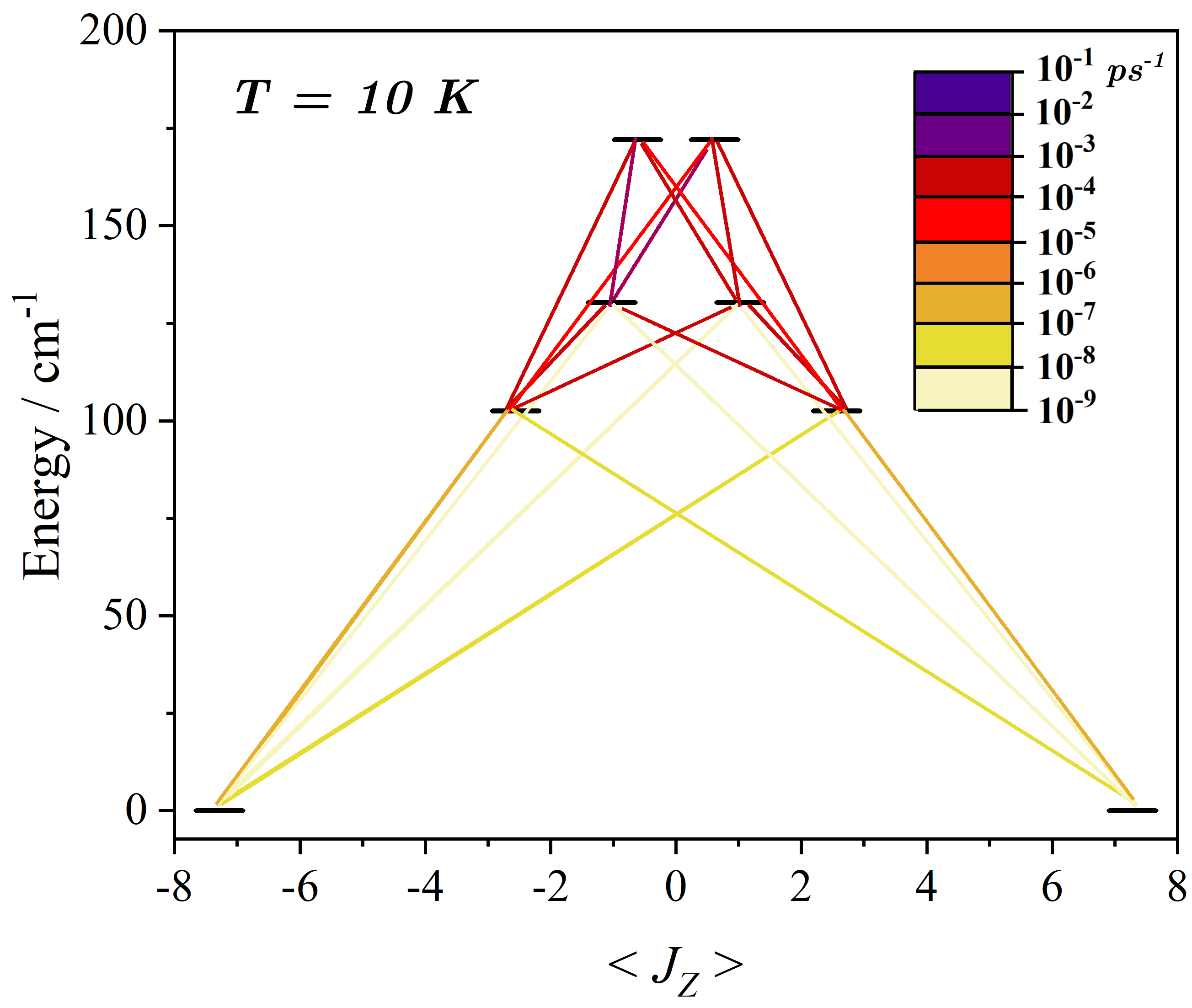}
  \hspace{0.5cm}
  \includegraphics[scale=0.35]{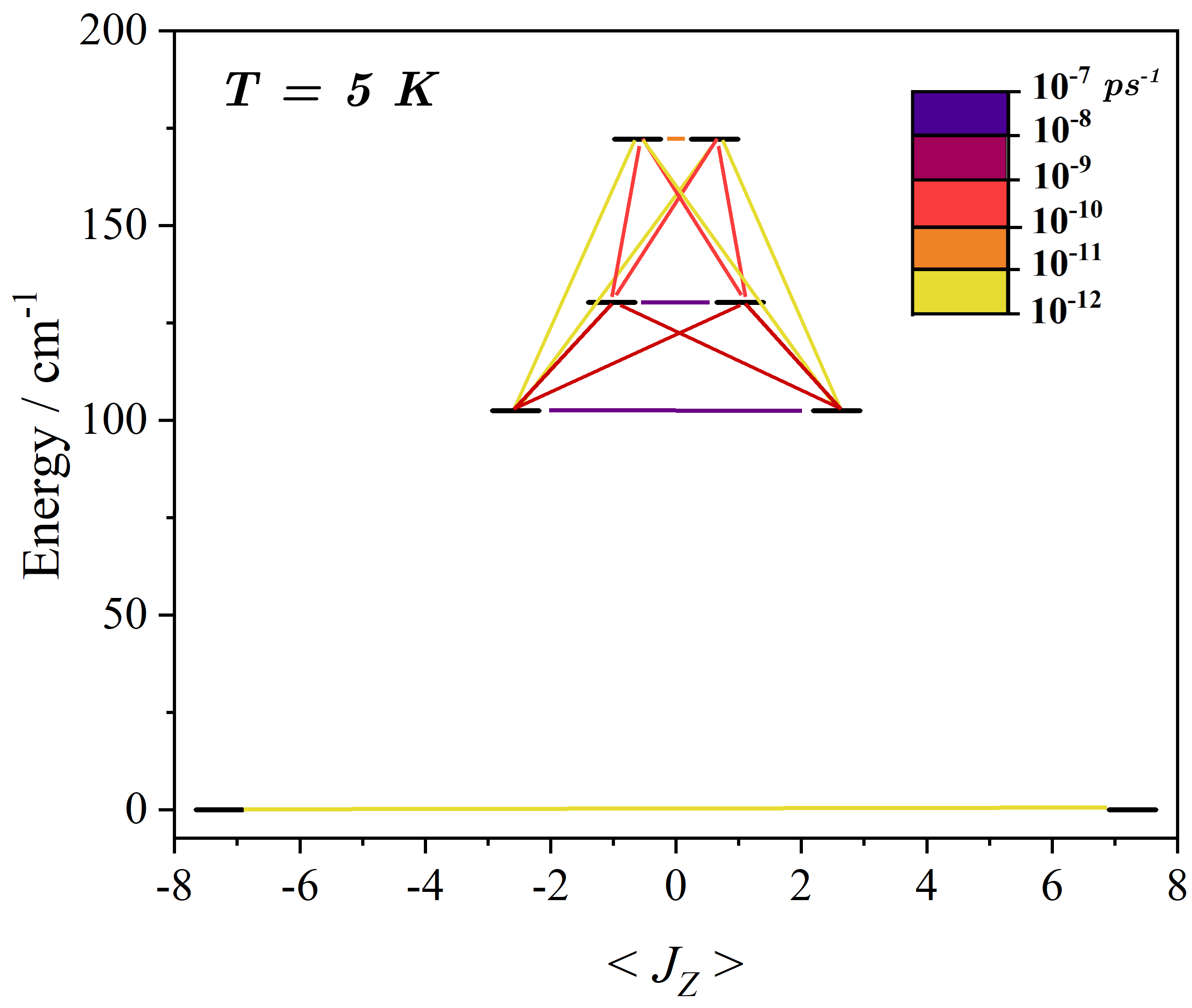}
 \end{center}
 \caption{\textbf{Finite temperature Orbach and Raman computed absorption rates.} Right and left panel report the matrix elements of $W^{1-ph}$ and $W^{2-ph}$, respectively, expressed in the basis of the eigenvectors of the ground KD's $g$-tensor. On the $x$-axis, it is reported the computed average magnetic moment for the first four KDs and their energy separation from the ground state. All the rates are expressed in $ps^{-1}$ and always refer to absorption rates. The emission rates are always found to be much larger because of the spontaneous emission contribution.}
 \label{Rates}
\end{figure*}

All parameters in Eqs. \ref{order1} and \ref{order2} are computed fully \textit{ab initio} with the sole exception of a Gaussian smearing used to approximating the Dirac delta function. The little dependency of the results on this parameter is discussed in SI. We also note that this degree of freedom can also be eliminated by further including the integration of the phonons across the Brillouin zone.\cite{Lunghi2019,Lunghi2020} The relaxation time $\tau$ for the Orbach and Raman relaxation is computed as the second-smallest eigenvalue of the matrices $W^{1-ph}$ and $W^{2-ph}$ and it is reported in Fig. \ref{Rates} for different values of temperature\cite{Sessoli}. 

\begin{figure}[htp]
 \begin{center}
  \includegraphics[scale=1]{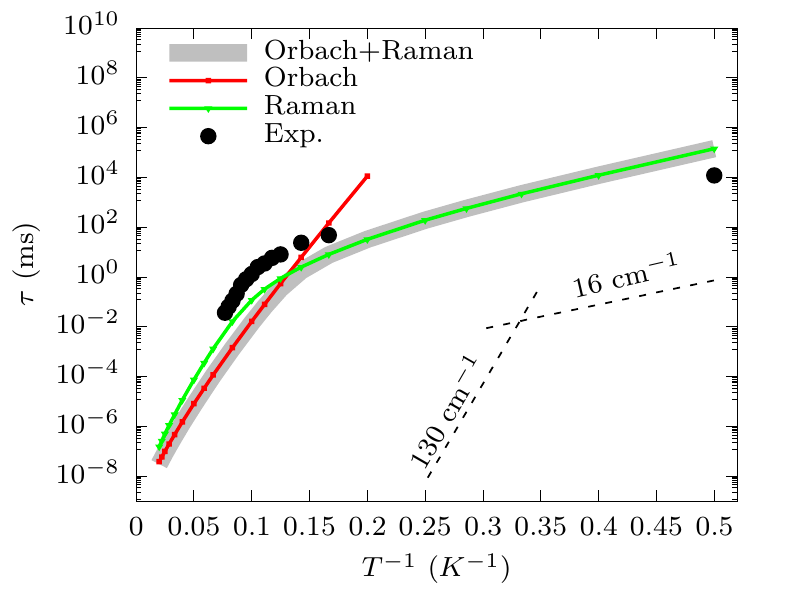}
 \end{center}
\caption{\textbf{Temperature Dependence of the Spin-Phonon Relaxation Time.} The continuous gray line reports the total computed relaxation time coming from both Orbach and Raman mechanism. The former contribution is reported explicitly with a red line and red squares, while the latter is reported with a green line and green triangles. The black circles reports the experimental relaxation times obtained from AC susceptibility measurements in absence of external field, for $6 K \le T\le 13 K$, and from hysterisis loop at 1500 Oe external field, for $T=2$ $K$. Dashed black lines corresponding to $exp(U/k_{B}T)$, with $U$=130 cm$^{-1}$ and 16 cm$^{-1}$ are reported in the bottom right part of the plot as a guide to the eye to interpret results.}
 \label{tau}
\end{figure}

The Orbach and Raman mechanisms are predicted to dominate relaxation at high and low temperature, respectively. The value of $U$ associated to the Orbach mechanism is generally extracted by fitting the slope of the linear relation of $ln(\tau)$ vs $(k_{B}T)^{-1}$. Simulations show that a perfect linearity is absent, but focusing on the high temperature data it is possible to identify a slope of $U\sim 130$ cm$^{-1}$. This value coincides with the energy of the second excited KDs. This is in agreement with the expression for the Orbach mechanism of Eq. \ref{order1}, where phonons in resonance with the electronic transitions are more efficiently absorbed or emitted. However, the non-perfect linearity of $ln(\tau)$ vs $(k_{B}T)^{-1}$ suggests that the relaxation is mediated by different KDs at different temperatures, with higher excited ones taking over relaxation as temperature increases and, at the same time, resonant phonons become populated. On the other hand, Raman mechanism is computed to dominate relaxation at low temperature, with the relaxation time following a power law with respect to temperature up to $\sim 8$ K. 

To get a deeper insight into the temperature dependence of the relaxation time, we analyzed the transition rates for the two mechanisms. The left panel of Fig. \ref{Rates} describes the computed transition rates for the Orbach mechanism at the temperature of 10 K. It should be noted that such transition rates are not computed as the expectation value of the magnetic dipole moment, as it is often reported in literature\cite{Blagg2013,Liu2016}, but instead are computed from the true \textit{ab initio} spin-phonon transition rates used to determine $\tau$ itself. Fig. \ref{Rates} highlights that an intra-ground KD transition has zero probability, as a consequence of the Kramers theorem, but the interwell transition from the ground state KD to the first and second excited KDs is very likely, and offers an efficient relaxation pathway. The transition rates of the Raman relaxation at 5 $K$ are reported in the right panel of Fig. \ref{Rates} and show a reverse situation with respect to the Orbach case. At the second order of perturbation theory the intra- ground-state KD transition is possible and dominates the relaxation mechanism. This is made possible by the presence of the excited KDs, which allows the simultaneous absorption and emission of two phonons. Interwell transitions among the excited states are also promoted by the Raman mechanism but have a negligible role at low temperatures because these levels are not populated.  

Eq. \ref{order2} shows that at the second order of perturbation theory, among all the phonons in the spectrum, those contributing to relaxation are those best satisfying three requirements: i) the absorbed phonon is more thermally populated than others, ii) its energy must be close to the transition energy of the excited KDs, iii) the energy conservation condition contained in $G^{2-ph}$ requires that the sum of the energy of absorbed and emitted phonons matches the energy difference between the initial and final spin state. In the case of a Kramers system in zero external field, condition iii) implies that the absorbed and emitted phonons must be degenerate. Considering Raman relaxation at 5 K, conditions i) and ii) are in antithesis, as only phonons with much lower energy of the first excited KD, computed at 100 cm$^{-1}$, will be significantly populated. Which of the two requirements is more stringent can be deducted by the fact that phonons' thermal population diminishes exponentially with their energy, as enforced by the thermal population term present in Eq. \ref{Gsph}, while requirement ii) is only enforced to the power of two, as evinced by the denominator of the relaxation rate prefactor in Eq. \ref{order2}. Therefore, the best compromise of conditions i)-iii) at low temperature is fulfilled by the absorption and emission of a degenerate pair of the first available optical modes, here computed to start at $\sim 16$ cm$^{-1}$. Such a normal mode is a collective rigid bending of all the carboxyl-Dy bonding angles (see Fig. S4) It affects the whole Ln coordination geometry but, at the same time, without distorting the ligand's carbon backbone (\textit{vide infra}).

Thus, for the Raman relaxation that directly connects the two sides of the anisotropy barrier in the regime of $\hbar\omega_{\alpha}>k_{B}T$, two degenerate phonons of this lowest in energy normal mode contribute to the $T$ dependence of $\tau^{-1}$ through $G^{2-ph}=n_{\alpha}(n_{\alpha}+1) \sim exp(-\beta \hbar\omega_{\alpha})$. Such a contribution would appear as straight line in the $log(\tau)$ vs $1/T$ plot of Fig. \ref{tau}, with slope of $\sim 16$ cm$^{-1}$. However, several pairs of phonons become more and more populated for increasing values of $T$, therefore adding their contribution to $\tau$. This translates into a power-law relation of $\tau$ vs $T$ in the low-$T$ Raman-dominated relaxation regime. 

Now that we have fully analysed the computational results, we turn to their comparison with experimental ones. Magnetization dynamics of this compound was previously investigated by Jiang \textit{et al.}\cite{Jiang2010}. However, given the tendency of the crystals to loose ethanol crystallization molecules, the characterization was repeated using a freshly synthesized \textbf{Dy$_{0.1}$acac} sample (90${\%}$ diamagnetically diluted sample in Y$^{3+}$ isostructural analogous) removed from the mother liquors immediately before the measurements (more details in Methods in SI). The relaxation times as a function of the inverse of temperature, shown in Fig. \ref{tau}, well compare with previous results. To acquire data on a larger temperature range, we extracted the relaxation time of magnetization by hysteresis measurements performed at 2 K (see Methods in SI). As it becomes clear comparing experiments and simulations, the range of temperatures scanned experimentally falls exactly in the regime where the Raman mechanism, dominating at low temperature, gives way to the Orbach one, which drives spin relaxation at high temperature. Besides being able to semi-quantitatively predict relaxation time, our simulation allows for a unambiguous determination of the underlying mechanism of relaxation.

Comparing the agreement between computed and measured $\tau$, the former are around two orders of magnitude shorter. This discrepancy is in line with the state-of-the-art in the field\cite{Lunghi2020} and it is probably due to deficiencies in the prediction of the many \textit{ab initio} parameters used to compute the relaxation time. For instance, the comparison of measured and predicted CTM shows only semi-quantitative agreement suggesting that a more accurate determination of the Ln's CF might improve the agreement between relaxation times. Similarly, phonons calculations are known to be affected by small inaccuracies due to the lack of anharmonic shifts or deficiencies in the description of the lattice's dispersion forces from DFT theory. Vibrational shifts in the order of 1-10 cm$^{-1}$ are generally observed between simulations and experiments\cite{Garlatti2020}, and a careful benchmark of these methods is a mandatory next step for the advancement of the field. 
Interestingly, the computed relaxation time at 2 K is slightly longer than the experimental one. This might result from the fact that experimental data are affected by a residual contribution of dipolar relaxation, that speeds up relaxation rates, and simulated relaxation times are computed without including low-energy acoustic phonons. It has been shown for a Co$^{2+}$ SIM that this approximation does not affect dramatically the results but might slightly reduce spin lifetime at very low temperature, where only the acoustic phonons are significantly populated.\cite{Lunghi2020}

Interestingly, the computed temperature dependence of $\tau$ due to the Raman process described here cannot be easily disentangled from the Orbach mechanism in the presence of anharmonic phonons\cite{Lunghi2017,Lunghi2020a}. A correct estimation of anharmonic interactions will require the computation of linewidths from first-principles and perturbation theory\cite{Califano}. Such a calculation is beyond the reach of current simulations and we postpone the pursuit of such a challenge to future work. However, we notice that some authors\cite{Yu2020,Reta2021} have recently reported that anharmonicity is unable to reproduce the experimental temperature dependence of $\tau$. As discussed in SI, such a discrepancy from our findings results from the assumption of a Gaussian smearing in place of the physically-correct Lorentzian one proposed in ref. \cite{Lunghi2017}, together with the lack of a proper integration of the vibrational Brillouin zone\cite{Lunghi2019, Lunghi2020, Lunghi2020a}.

\section*{The Role of Molecular Vibrations and Electrostatic Polarization}

\begin{figure*}
 \begin{center}
  \includegraphics[scale=1]{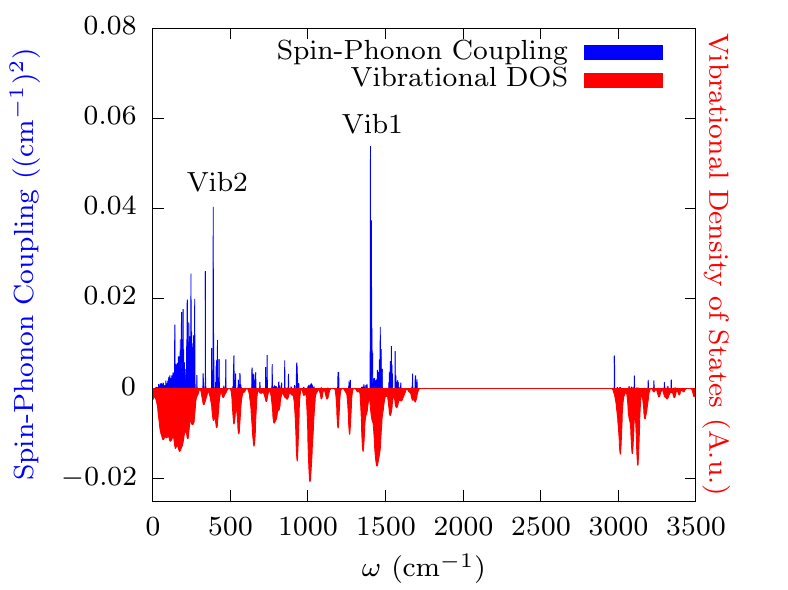}
  \includegraphics[scale=1]{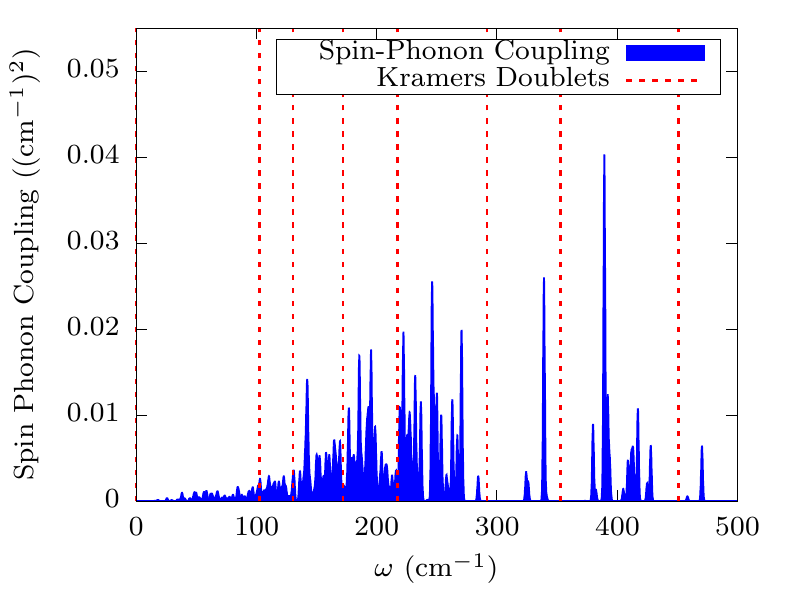}
 \end{center}
\caption{\textbf{Spin-Phonon Coupling Distribution and Vibrational Density of States.} The left panel reports the spin-phonon coupling distribution and the vibrational density of states as function of energy. A Gaussian smearing with $\sigma=1$ cm$^{-1}$ and $\sigma=10$ cm$^{-1}$ has been applied to the two functions, respectively. The right panel is a close-up of the spin-phonon coupling distribution overlapped to the energy resonance of the KDs.}
 \label{sph}
\end{figure*}

Having identified how one- and two- phonons mechanisms affect the temperature dependence of the relaxation time, we can now turn to the analysis of spin-phonon coupling and how it is influenced by the chemical structure and composition of \textbf{Dyacac}'s lattice. The left panel of Fig. \ref{sph} reports the average spin-phonon coupling coefficients as a function of the phonon's frequency, together with the vibrational DOS. The right panel of Fig. \ref{sph} instead shows more in details the spin-phonon coupling in the spectral region that overlaps with the KDs energies. No strict correlation between the spin-phonon coupling intensity and the vibrational density of states emerges, and this is a consequence of the dependence of spin-phonon coupling strength on the nature of the specific vibration. For instance, spectral region around 1000 cm$^{-1}$ is significantly populated by vibrations but they do not show any significant coupling to the spin. A visual inspection of these modes reveals that the coordination sphere of \textbf{Dyacac} is only slightly involved in these vibrations, which are instead strongly localized on the acacs' methyl groups rotations or entirely on other molecules, such as on the EtOH molecules. The same effect is observed for the lowest part of the spectrum. Starting from zero energy, spin-phonon coupling increases at a much slower rate than the vibrational density of states. This is due to the fact that the lowest energy modes are dominated by rigid translations of the molecules in the lattice. Such kind of motions are not in principles coupled to spin but their slight admixing with intra-molecular distortions can lead to a finite coupling\cite{Lunghi2017a,Lunghi2019}. Such an admixture (Fig \ref{sph}, right panel) increases as the energy of vibrations approaches the typical values of intra-molecular modes in the THz spectral region.

\begin{figure*}[ht]
 \begin{center}
  \includegraphics[scale=0.7]{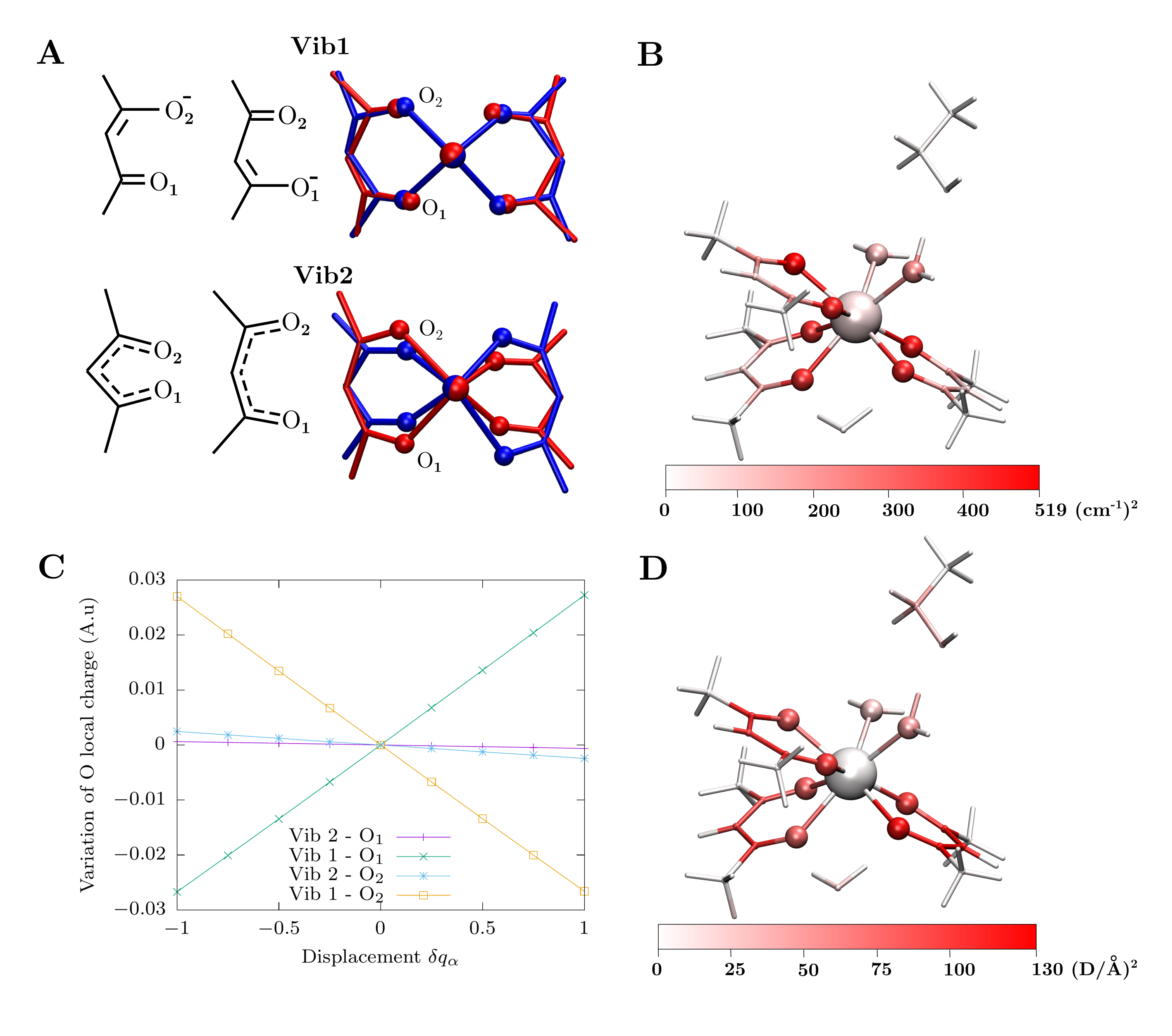}
  \end{center}
\caption{\textbf{Electrostatic Contributions to Spin-Phonon Coupling.} \textbf{A}. Schematic representation of \textbf{vib1} and \textbf{vib2}. For the sake of clarity we showed only the two acac ligands where most parts of the normal modes are localized (acac1 and acac2 in Fig. S1). \textbf{B}. Magnitude of the spin-phonon coupling coefficients resolved by atomic site. \textbf{C} Variation of computed Loprop atomic charge on O$_1$ and O$_2$ as a function of the displacement along the normal coordinates for \textbf{vib1} and \textbf{vib2}. \textbf{D}. Magnitude of the first derivative of electric dipole resolved by atomic site. Module of the derivative for Dy has been set to zero (see SI) to highlight the values' variations in the organic scaffold.}
 \label{Atoms}
\end{figure*}

Based on these observations and previous studies, one would be tempted to conclude that the largest spin-phonon coupling is triggered by molecular motions strongly distorting the first coordination shell of Dy$^{3+}$. Although not wrong, the analysis of \textbf{Dyacac}'s spin-phonon coupling reveals that this is not the entire picture. We now perform a detailed analysis for two modes among the most strongly coupled ones: \textbf{vib1} and \textbf{vib2}, with resonance at $\sim$ 1500 cm$^{-1}$ and $\sim$ 380 cm$^{-1}$, respectively. Both modes are labelled in the left panel of Fig. \ref{sph} and they are pictorially shown in Fig. \ref{Atoms}A. The choice of these two modes resides on the fact that they show a very high coupling and at the same time present rather different features in terms of atomic displacements. \textbf{vib2} is characterized by a significant distortion of the first coordination shell, by breathing of the Dy-acac chelate rings, while carbonyl CO bond lengths remain quite rigid. \textbf{vib1} is largely characterized by a stretching of all the acac ligand's carbonyl groups, with Dy-O distances being less affected. \textbf{vib1}, differently from \textbf{vib2}, is expected to significantly affects the acac's conjugated system by localizing the $\pi$ electrons towards the two limit Lewis configurations. We note that these vibrations mostly involve two acac ligands, leaving the third one and the water molecules almost unaffected. Although there are other vibrations of similar nature but affecting other pairs of ligands, this does not limit the generality of our discussion.

The nature of \textbf{vib1} suggests that atoms beyond the first coordination shell might play a significant role in inducing relaxation. However, normal modes are always delocalized on the entire molecules to some degrees, and even \textbf{vib1} presents some distortions of the distances and angles between the Dy$^{3+}$ ion and oxygen atoms. We computed the magnitude of the spin-phonon coupling coefficients resolved by atomic site, as defined in the Methods sections in SI. Results are plotted in Fig. \ref{Atoms}B, and show a remarkable finding: the sp$^{2}$ carbon atoms belonging to the acac ligands, namely the Dy's 3rd nearest neighbours, are found to be coupled to spin at least as much as the oxygen atoms of the waters directly bound to Dy. In order to understand the origin of this phenomenon we perform a detailed analysis of the role of covalent and electrostatic interaction on spin-phonon coupling. 

For both \textbf{vib1} and \textbf{vib2} we compute an atomically-resolved expansion of the electrostatic potential for different vibrational amplitudes and compute the crystal field parameters by replacing one acac ligand (acac1 in Fig S1) with the corresponding atomic point charges, dipoles and quadrupoles\cite{Gagliardi2004}. The spin-phonon coupling coefficients obtained with an electrostatic model that includes only the contribution of the acac's oxygen donor atoms are in excellent agreement with the original ones obtained with the explicit model. This demonstrates that, although the static crystal field is generally determined by both electrostatic\cite{Sorace2011} and covalent interactions\cite{Briganti2019}, its modulation by small atomic displacements, \textit{i.e.} the spin-phonon coupling, is largely driven by the electrostatic effects of \textbf{Dyacac}'s first coordination sphere.

In order to reconciliate the observation of a strong effect of Dy's 2nd and 3rd nearest neighbours and a spin-phonon coupling dominated by the sole electrostatic contribution of the first coordination shell, we advance the hypothesis that \textbf{Dyacac}'s second coordination sphere acts indirectly through a polarization of the first coordination shell's local charge. In other words, the acac $\pi$ system can allow a strong polarization of the CO bonds.  This idea is supported by fact that \textbf{vib2}, being characterized by a rigid motion of the ligands's intra-molecular bonds, do not modulate significantly the local charge distribution of the first coordination shell, while the carbonyl stretching motions characterizing \textbf{vib1} lead to local charges fluctuations two orders of magnitude larger (see Fig. \ref{Atoms}C). 

The smoking gun of the role of charge polarization is provided by repeating the calculation of the spin-phonon coupling for \textbf{vib1} and \textbf{vib2} by fixing the electrostatic parameters to those of the equilibrium geometry. In the case of \textbf{vib2} no difference is observed with respect to the calculation done adapting the charges during the vibrational motion, while large deviations are observed for \textbf{vib1}. Interestingly, by removing the charge polarization effect in \textbf{vib1}, the results become closer to those obtained by removing the acac ligand altogether (see Fig. S6). Additional evidence of the correlation between electrostatic polarization and spin-phonon coupling is provided by the computation of the effect of atomic displacements on the molecular dipole moment. The results are plotted in  Fig. \ref{Atoms}D, in the same way as for the atomically resolved spin-phonon coupling. With exclusion of the Dy atom, the two plots correlate very nicely, further supporting the electrostatic mechanism for non-local spin-phonon coupling.

It is important to remark that although all these considerations do not take into account the thermal population of vibrations and the resonance conditions imposed by Eqs. \ref{order1} and \ref{order2}, the discussion of \textbf{vib1} and \textbf{vib2} is crucial to highlight the origin of spin-phonon coupling in Ln ions beyond the simple argument of the relevance of distortions affecting the first coordination shell. Moreover, the spin-phonon coupling intensity of high-energy vibrations as the ones discussed here, would determine the $\tau_{0}$ of the Orbach process of molecules showing high axiality and large splitting of the electronic levels. Although this is not relevant for explaining the relaxation behaviour of \textbf{Dyacac} (\textit{vide infra}), it becomes of crucial importance for the design of new high-performance Ln SIMs with values of $U$ above 1000 cm$^{-1}$.

Finally, we turn to the discussion of the modes directly involved in the spin-lattice relaxation of \textbf{Dyacac} due to the constraints introduced by thermal population and resonance conditions. The latter is evidenced in the right panel of Fig. \ref{sph}, where the energy of the computed spin levels are superimposed to the spin-phonon coupling diagram of the low energy vibrations. The modes close in energy to the second excited KDs, responsible for high-temperature relaxation, are rather delocalized on the molecular unit, but it is still possible to discern a significant contribution of ligand's methyl groups rotations and torsions of the entire acac's sp$^{2}$ structure. These motions also accompany a modulation of the Dy-O bonds' length and O$\widehat{\textrm{Dy}}$O angles. Raman relaxation receives contributions from a large number of vibrations, however, at low temperature, only the first few modes are significantly populated and able to contribute. As discussed before these modes are delocalized on the entire unit cell and are characterized by almost-rigid translations of the molecules in space overlapped to slight rigid rotations and intra-molecular distortions. We note that the acacs' Me group rotation are often involved in these mixed modes. This has been also observed for molecular qubits presenting acac ligands\cite{Lunghi2020,Garlatti2020} suggesting a critical role of this low-energy motion in promoting admixture of intra- and inter-molecular motions. 

\section*{Discussion and Conclusions}

The advent of SIMs with very large zero-field-splitting has led to a strong deviation of spin relaxation from the expected Orbach trend at low temperature. Since then, a Raman relaxation mechanism has been postulated to take place in that regime and to follow a phenomenological power-law $\tau^{-1} \propto T^{n}$, with $2<n<6$, regardless the standard picture of Raman relaxation theory predicts $n=9$ for Kramers systems\cite{Abragam}. Such a phenomenological approach however hides the underlying physical process responsible for relaxation and does not provide a clear indication on how we can to tackle the problem of fast Raman relaxation at low temperature. It has now been clearly demonstrated that the low energy vibrational DOS of molecular crystals significantly deviates from the Debye model and that optical modes appear at surprisingly low energies. Under these circumstances, the temperature dependence of Raman relaxation across degenerate KDs is shown to follow the Fourier transform of the two-phonon correlation function of pairs of degenerate phonons that are simultaneously absorbed and emitted from the low-energy vibrational spectrum, each contributing to $\tau^{-1}$ with a factor
\begin{equation}
    \bar{n}_{\alpha}(\bar{n}_{\alpha}+1)=\frac{e^{\hbar\omega_{\alpha}/k_{B}T}}{(e^{\hbar\omega_{\alpha}/k_{B}T}-1)^{2}}
    \label{Ramantau}
\end{equation}
In the limit of $k_{B}T<\hbar\omega_{\alpha}$, Eq. \ref{Ramantau} reduces to the exponential expression $\tau\sim exp(\hbar\omega_{\alpha}/k_{B}T)$, where $\hbar\omega_{\alpha}$ is of the order of the first $\Gamma$-point optical modes. Summing up contributions from different phonons at different temperatures, Eq. \ref{Ramantau} leads to the $\tau^{-1} \propto T^{n}$, with $2<n<9$.

Interestingly, in the limit of $k_{B}T>\hbar\omega_{\alpha}$, Eq. \ref{Ramantau} leads to $\tau\sim T^{-2}$, which has been recently shown to dominate high-temperature relaxation in $S=1/2$ molecular qubits\cite{Lunghi2020a}. It must be noted, however, that even though the phonon contribution to Raman relaxation in $S=1/2$ and high-spin SIMs follows the same $T$-dependence, the origin of spin-phonon coupling is qualitatively different. For $S=1/2$ systems in external field or in the presence of dipolar/hyperfine fields, Kramers theorem does not prevent the transition between the $m_{S}=\pm 1/2$ states, and Raman relaxation is due to the quadratic dependence of the spin Hamiltonian with respect to atomic displacements\cite{Lunghi2020a,Eaton2001}. Instead, zero-field Raman relaxation in high-spin Kramers SIMs depends on the linear term in the spin-phonon coupling with respect to the phonon variables, but at the second order in time-dependent perturbation theory, which is made possible by low-lying excited KDs states, as depicted by Eq. \ref{order2}. The identification of the correct relaxation mechanism is not only crucial for the correct interpretation of relaxation times but it also provides a clear pathways to understanding how spin-phonon coupling arises from molecular motions and electronic structure. Indeed, in this contribution it has been possible to analyze the spin-phonon coupling on a single-vibration basis and reveal that both modulations of the geometry of the first coordination sphere, as well as electrostatic polarization effects are at the origin of the coupling of the spin with the lattice.

On the basis of these new findings it is finally possible to draw an updated list of requirements for disengaging the magnetic molecular degrees of freedom from the vibrational thermal bath. Many of the guidelines that will follow have already been suggested by others in the context of slowing down the Orbach mechanism or the magnetization tunneling and here we will provide a comprehensive view of their effect on Raman relaxation. There are three different aspects that pertain to spin-phonon relaxation: \textbf{1)} the nature and symmetry of the static crystal field, \textbf{2)} the population of phonons participating to relaxation, and \textbf{3)} their spin-phonon coupling value. Each of these three aspects can be acted upon by implementing the following guidelines:

\begin{itemize}
    \item[\textbf{1a.}] Kramers ions with high $J$ values\cite{winpenny2013,Ungur2016};
    \item[\textbf{1b.}] Large crystal field splitting of the ground $J$ multiplet\cite{Ungur2011,Rinehart2011};
    \item[\textbf{1c.}] Strong axiality of the entire crystal field, not limited to the $g$-tensor of the ground state KD \cite{Ungur2011,Chilton2015b,Bonde2020}; 
\end{itemize}    

\begin{itemize}    
    \item[\textbf{2a.}] Vibrational density of states staggered from spin resonances\cite{Ullah2019};
    \item[\textbf{2b.}] Small amount of low-energy vibrations in both the lattice and the molecular unit\cite{Lunghi2017,Lunghi2017a,Chiesa2020};
\end{itemize}

\begin{itemize}    
    \item[\textbf{3a.}] Small admixing of intra- and inter-molecular vibrations at low energy\cite{Lunghi2017a,Lunghi2019};
    \item[\textbf{3b.}] Use of ligands with donor atoms' local charge not affected by local vibrations.
\end{itemize}

Although all these rules apply to both Orbach and Raman relaxation, the efficiency of their implementation will generally be different for the two mechanisms. For instance, enforcing the points 1b and 1c will increase the value of the KD doublets mediating relaxation. The latter quantity is directly linked to the energy of the phonon mediating Orbach relaxation, thus influencing relaxation rate exponentially, but it only enters to the power of two in the definition of the Raman rate prefactor (see Eqs. \ref{order1} and \ref{order2}). As a consequence, if increasing relaxation time at low temperature is the goal, points 1b and 1c are probably not sufficient because the Raman process will still dominate the relaxation. This is also a consequence of the fact that the energy of the phonons contributing to Raman relaxation is not necessarily in resonance with the spin states and low-energy vibrations will always dominates the two-phonon relaxation. Raman relaxation, however, will be more sensitive to the guidelines 3a. and 3b. These measures act to reduce the effective coupling of spin to low-energy vibrations, which enters to the fourth power into the relaxation rate expression instead of to the second in the case of Orbach. 

To the best of our knowledge, guideline 3b has never been reported before. As a rule-of-thumb, we can state that the electrostatic polarization of the donor atom with respect to the vibrational modes strongly affects the coupling with the spin. Moreover, we expect that such phenomenon is enhanced for ligands where multiple resonance limit formulae involving the donor atoms can be drawn. Indeed, conjugated sp$^2$ bonds also allow atoms beyond the second-coordination sphere to efficiently modulate the donor's charge in virtue of electronic delocalization, and therefore contribute to spin-phonon coupling. These compounds are more prone to have significant spin-phonon couplings of vibrations connecting first and n$^{th}$ coordination sphere (see SI for a comparison among acac, acetone and 2-propanol, Fig. S12). Such a guideline provides further emphasis to the importance of exploring organometallic compounds based on haptic ligands, which might combine high rigidity (2b,3a) and low level of atomic polarizability (3b) due to local molecular distortions induced by vibrational activity. In this last case the resonance formulae that can be drawn are equivalent from the point of view of charge delocalization on the donors and, as a consequence, the relaxation induced by charge modulation should not play a major role. We expect such a strategy to be already accounted for in the class of dysprosocenium single molecule magnets\cite{Goodwin2017,Guo2018} and similarly designed molecular spin qubits\cite{Ariciu2019,DeCamargo2020}. Finally, we remark that these considerations are likely to affect more strongly the high-temperature Orbach regime of SIMs with large $U$, where relaxation proceed through the absorption of phonons with energy comparable to the stretching of rigid bonds such as those active in \textbf{vib1}.

In conclusion, we have applied \textit{ab initio} spin dynamics to the case of a Dy-SIM and successfully explained the origin of its slow spin relaxation. Our method provided the first successful prediction of both one- and two-phonon relaxation rates in a Ln compound free from any adjustable parameter and allowed to completely bypass phenomenological approaches commonly used in the field. The identification of the correct mechanism of relaxation made it possible to derive an updated list of requirements to slow down both Orbach and Raman relaxation. In particular, from the analysis of \textit{ab initio} vibrations and spin-phonon coupling coefficients, it was possible to derive new insights on the molecular quantities that regulate relaxation rates across all temperature ranges. Of particular relevance is the finding that electrostatic polarization plays an active role in determining spin-phonon coupling, offering a new chemically-sound pathway to control spin relaxation. 

\vspace{0.2cm}
\noindent
\textbf{Data Availability}\\
All the relevant data discussed in the present paper are available from the authors upon request. \\

\vspace{0.2cm}
\noindent
\textbf{Acknowledgements}\\
This project has received funding from the European Research Council (ERC) under the European Union’s Horizon 2020 research and innovation programme (grant agreement No. [948493]) and from AMBER (grant 12/RC/2278\_P2). Computational resources were provided by the Trinity Centre for High Performance Computing (TCHPC) and the Irish Centre for High-End Computing (ICHEC). We acknowledge Prof. Lorenzo Sorace and Dr Mauro Perfetti for critically reading the manuscript. The authors acknowledge Prof. Fernando Luis for the help in the analysis of hysteresis results. \\

\vspace{0.2cm}
\noindent
\textbf{Present Affiliations}\\
Dr Lorenzo Tesi's present affiliation is \textit{Institute of Physical Chemistry, University of Stuttgart, 70569 Stuttgart, Germany} \\

\vspace{0.2cm}
\noindent
\textbf{Conflict of interests}\\
The authors declare that they have no competing interests.

%


\end{document}